\documentstyle[12pt,psfig]{article}
 
\begin{document} 
\begin{titlepage} 
\begin{center} 
{\large {\bf Susceptibility amplitude ratio \\ 
in the two-dimensional three-state Potts model}} \\ 
\vspace{1.5cm} 
{\bf L.~Shchur$^{a,b,c}$, P.~Butera$^{b}$ and B.~Berche$^{c}$} \\ 
\vspace{0.8cm} 
 
$^a${\em Landau Institute for Theoretical Physics, 142432 Chernogolovka, 
Russia} \\ 
 
$^b${\em Instituto Nazionale di Fisica Nucleare, Universit\'a Milano-Bicocca, 
Piazza delle  Scienze 3, 20126, Milano, Italia} \\ 
 
$^c${\em Universit\'e Henri Poincar\'e - Nancy I, Laboratoire de Physique 
des Mat\'eriaux, B.P. 239, 54506 Vand\oe uvre les Nancy Cedex, France } 
\end{center} 
\vspace{6mm} 
 
\begin{abstract} 
\noindent 
 
We analyze Monte Carlo simulation and series-expansion data for the
susceptibility of the three-state Potts model in the critical region. The
amplitudes of the susceptibility on the high- and the low-temperature
sides of the critical point as extracted from the Monte Carlo data are in
good agreement with those obtained from the series expansions and their
(universal) ratio compares quite well with a recent quantum field theory
prediction by Delfino and Cardy.
 
\end{abstract} 
\end{titlepage} 
 
\newpage 
 
\section{Introduction} 
 
The universal thermodynamic behaviour of a system in the vicinity of a
critical point is characterized by a set of critical exponents and by
universal combinations of critical amplitudes~\cite{PHA-rev}.  While the
critical exponents are generally well studied, there are still few
theoretical results on the universal combinations of critical amplitudes.
Recently, some progress was achieved by Delfino and Cardy~\cite{DC} (this
reference shall be denoted as article~I throughout our paper) for the
two-dimensional $q$-state Potts model and some universal amplitude ratios
were computed for $q=2,3$ and $4$ (see also ~\cite{M} for the most recent
review on other results.)
 
It is commonly accepted that for a typical spin model, for instance the
Ising model, there are only two independent length scales~\cite{SFW} in
the critical region and thus there must be four universal
relations~\cite{AH} among the six following critical amplitudes: the
amplitudes of specific heat in the ordered phase $A_-$ and in the
symmetric phase $A_+$; the analogous amplitudes of the magnetic
susceptibility $\Gamma_-$ and $\Gamma_+$; the amplitude of the
magnetization when approaching the critical temperature from below,$B$;
and the amplitude of the dependence of the magnetization on the magnetic
field at the transition temperature, $D$. In the case of the
two-dimensional Potts model with $q >2$, it is possible to define also a
``transverse susceptibility'' in the low temperature phase~\cite{DBC}. In
terms of its critical amplitude $\Gamma_T$, a fifth universal ratio
$\Gamma_T/\Gamma_-$ can be defined which is determined only by the
behaviour in the LT phase.

The values of the ratio $\Gamma_+/\Gamma_-$ of the susceptibility critical
amplitudes, were calculated in paper~I with the results $37.699$, $13.848$
and $4.013$ for $q=2,3$ and $4$, respectively. The first value coincides
with the well known exact result~\cite{WMTB} in the four digits presented.
In Ref.~\cite{DBC} Delfino, Barkema and Cardy performed a Monte Carlo (MC)  
test of the other predictions of paper~I and found results for
$\Gamma_+/\Gamma_-$ not consistent with their expectations in the $q=3$
case, while in the $q=4$ case their results were inconclusive.  Actually
the analysis of the data in the $q=4$ case is somewhat difficult due to
the expected logarithmic corrections~\cite{NS,CNS,SS} to the power-like
critical behaviour and the results~\cite{CTV} of a multi-parameter fit are
still controversial (see, for instance Ref.  \cite{DBC}).  On the other
hand their analytical calculations for $\Gamma_T/\Gamma_-$ are in very
good agreement with the numerical results reported in Ref.~\cite{DBC}.
 
Here we present an analysis of the susceptibility of the $q=3$ Potts model
by a MC simulation supported by extrapolations of the presently available
LT and HT series expansions. The results of both procedure are completely
consistent in the critical region window. The simplest extrapolation
 of the series expansions by Pad\'e approximants yields the estimate
$\Gamma_+/\Gamma_-=14.2(3)$, while a fit of the MC data leads to
$\Gamma_+/\Gamma_-=14\pm 1$. Thus the agreement with the Delfino and Cardy
prediction for $q=3$ is very good. Since also their estimate of the
susceptibility ratio for $q=2$ is correct, our result gives us further
confidence that even the prediction for the $q=4$ case might be correct.  
However, further numerical study of this case would still be welcome.
 
The text is organised as follows. In section~\ref{Model} we define the
model and present a few details of the series calculations and of the MC
simulations. The data fitting procedure is discussed in the
section~\ref{Results} and the results are summarised and commented in
section~\ref{Summary}.
 
\section{Model and Definitions} 
\label{Model} 
 
The Hamiltonian of the Potts model is 
 
\begin{equation} 
H=\sum_{<ij>}(1-\delta_{s_i,s_j})+h \sum_i(1-\delta_{s_i,0}) 
\label{Ham1} 
\end{equation} 
 
\noindent where $s_i$ is a site variable taking $q$ integer values between 
$0$ and $q-1$, and $h$ is an external magnetic field which stabilizes
the state $0$. 
 
The susceptibility is defined in terms of the free energy per site 
$f(\beta,h)$, where $\beta=1/T$ is the inverse temperature variable, 
starting with a finite system of $N$ sites and then taking the thermodynamic 
limit: 
 
\begin{equation} 
f(\beta,h)=\lim_{N\rightarrow\infty}-\frac{ 1} {N\beta} \log (Z_N(\beta,h)) 
\label{def-free} 
\end{equation} 
 
\noindent with a partition function $Z_N$ defined according to 
 
\begin{equation} 
Z_N(\beta,h)=\sum_{conf} e^{-\beta H}. 
\label{def-part} 
\end{equation} 
 
\noindent The susceptibility at inverse temperature $\beta$ is thus
 
\begin{equation} 
\chi(\beta)=\left. -\frac1\beta \frac{\partial^2f(\beta,h)}{\partial h^2} 
\right|_{h=0}\; . 
\label{def-susc} 
\end{equation} 
 
\subsection{Low-temperature and high-temperature expansions} 
 
The first few LT expansion coefficients can be very simply obtained by 
classifying the spin configurations with respect to their energy and 
multiplicity.  In the LT variable $z=e^{-\beta}$ we have 
 
\begin{eqnarray} 
\chi = (q-1)z^4+2(q-1)z^6+2(q-1)(q-2)z^7+... 
\label{x2} 
\end{eqnarray} 
 
It is by far less trivial to derive the very long LT expansions through 
$z^{47}$ tabulated in Ref.~\cite{BEG} for zero field in the $q=2,3$ and 
$4$ cases. 
 
On the other hand, the presently available HT expansions for the 
susceptibility are still of moderate length. For general $q$, they reach 
~\cite{SR} the order $10$ in the HT variable  
 
$$v=\frac{1-z}{1+(q-1)z}.$$ 
 
\noindent 
The first few terms of the expansion for $\chi$ are 
 
$$ 
\chi=1+4v+12v^2+36v^3+ (76+12q)v^4+... 
$$ 
 
 The HT series is however sufficient to compute accurately $\chi$ provided
that one does not get too close to the critical point
$\beta_c=\log(1+\sqrt{q})$.  It should be noted that the HT expansion
tabulated in Ref.~\cite{SR} is conventionally normalized to 1 at infinite
temperature and therefore it should be multiplied by the factor $\frac
{q-1} {q^2}$ to get the correct amplitude.
 
We recall that the critical exponent is given by the exact formula 
$$ 
\gamma(q) = \frac{7 \pi^2- 8\mu \pi +4\mu^2}{6\pi(\pi -2\mu)} 
$$ 
with $\sqrt q= 2\cos\mu$. 
 
 In terms of the HT series and of the known values of $\gamma$ and
$\beta_c$, we can form, on the HT side of the critical point, the ``HT
effective amplitude'' $\Gamma_+(\tau)= \tau^{\gamma} \chi(\tau)$. Here
$\tau = 1- \beta/\beta_c$ is the reduced inverse temperature. Similarly,
we can construct a ``LT effective amplitude'' $\Gamma_-(\tau)=
(-\tau)^{\gamma} \chi(\tau)$, on the LT side $\tau<0$ of the critical
point. The critical amplitudes $\Gamma_+$ and $\Gamma_-$ are computed by
extrapolating to the critical point the corresponding effective
amplitudes. Since the HT series is not very long, the extrapolation can
only be performed, in the most naive way, by Pad\'e approximants, which
cannot allow for the singular corrections to scaling. Therefore we should
not expect an accuracy of the amplitude $\Gamma_+$ better than a few
percent.  The LT series is much longer, but we have preferred to use
Pad\'e approximants also in this case. We have used the highest order
available approximants to extrapolate the effective amplitudes, namely the
[5/5] approximant in the HT region, leading to $\Gamma_+=0.180(3)$ and the
[22/22] approximant in the LT region, which gives $\Gamma_-=0.0127(1)$. In
both cases we have checked the accuracy of the results by comparing these
approximants with the lower order ones which are nearby in the Pad\'e
table. From this study we can conclude that $\Gamma_+/\Gamma_-=14.2(3)$.
Since the largest contribution to the error comes from the determination
of $\Gamma_+$, in order to improve the result, significantly longer HT
series should be computed and the analysis should be performed by
differential approximants, which can allow for the corrections to scaling.
 
\subsection{Monte Carlo simulations} 
 
We adopt the Wolff algorithm~\cite{W} for studying square lattices of
linear size $L$ with periodic boundary conditions. Starting from a typical
ordered state, we let the system relax in $10^4$ steps measured by the
number of flipped Wolff clusters. The averages are computed over $10^5$
steps. Our random numbers are produced by an exclusive-XOR combination of
two shift-register generators with the taps (9689,471) and (4423,1393),
which are known~\cite{S} to be safe for the Wolff algorithm.
 
The maximal system size $L=200$ we used is rather moderate for the current 
computer standards, but is quite sufficient for our purpose, as we will 
see. 
 
During the simulations we have evaluated an order parameter $M(t)$ defined
as follows
 
\begin{equation} 
M=\frac{q\frac{N_m}{N}-1}{q-1} 
\label{Order-Potts} 
\end{equation} 
 
\noindent where $N_m$ is the number of sites $i$ with $s_i=m$ at the time
$t$ of the simulation~\cite{B}, and $m\in [0...(q-1)]$ is the value of the
majority of the spins at that time. This is like using the modulus of the
magnetization in a MC simulations of Ising model.
 
Thus, the susceptibility in the LT phase is given by the fluctuations of 
the majority of the spins 
 
\begin{equation} 
k_BT\chi_-=\frac1N (<N_m^2>-<N_m>^2) 
\label{susc-LT} 
\end{equation} 
 
\noindent while in the HT  phase $\chi$ is given by fluctuations in all $q$ 
states, 
 
\begin{equation} 
k_BT\chi_+=\sum_{\mu=0}^{q-1}\frac{1}{qN} (<N_\mu^2>-<N_\mu>^2), 
\label{susc-HT} 
\end{equation} 
 
\noindent where $N_i$ is the number of sites $i$ with the spin in the
state $\mu$. Properly allowing for the finite-size effects, this
definition of the susceptibility gives, in both phases, an extremely good
consistency with the series expansion data.
 
\section{Data analysis} 
\label{Results} 
 
In this section we shall argue that system sizes such that $L\times L
\approx 100 \times 100$ sites are sufficiently large to reproduce the
susceptibility behaviour of the infinite system in the range of the
``critical temperature window''. Then we shall present a comparison of MC
and series data.  Finally, we shall discuss the fit to the data and
comment on the quality of the results.
 
\subsection{Critical region window} 
 
The critical region window $|\tau| \in [\tau_{fs},\tau_{corr}]$ is usually
characterized~\cite{TS,DBC} as an interval between two (inverse reduced)  
temperatures $\tau_{fs}$ and $\tau_{corr}$ in which the finite system of
size $L$ exhibits the critical behaviour of the infinite system. For
instance, in the case of the susceptibility we should have
 
\begin{equation} 
\chi \propto \Gamma_\pm |\tau|^{-\gamma} 
\label{chi-crit} 
\end{equation} 
 
\noindent where $\Gamma_\pm$ are the critical amplitudes in the HT phase
(+) and in the LT phase (-), respectively, and $\gamma$ is the critical
exponent. The ``left-hand'' boundary of the window $\tau_{fs}$ is the
rather well defined temperature at which the correlation length
$\xi\propto\tau^{-\nu}$ is of the order of the lattice size $L$, so
 that $\tau_{fs}\approx L^{-1/\nu}$. The ``right-hand'' limiting value of 
the critical window $\tau_{corr}$ is related to the corrections to scaling 
and could be defined as the temperature at which the deviation from the 
critical behaviour (\ref{chi-crit}) due to the corrections to scaling 
reaches the level of e.g. $ 1\%$. Therefore the temperature $\tau_{corr}$ 
could be well identified knowing exactly the corrections to scaling as in 
the case of the Ising model~\cite{TS,SV}. In the case of insufficient 
theoretical knowledge of the amplitudes of the correction terms, as is the 
case of the $3$-state Potts model, we can try to identify the 
``right-hand'' temperature boundary $\tau_{corr}$ by plotting the MC data 
for the previously defined ``effective amplitude''. 
 
\begin{figure} 
\psfig{figure=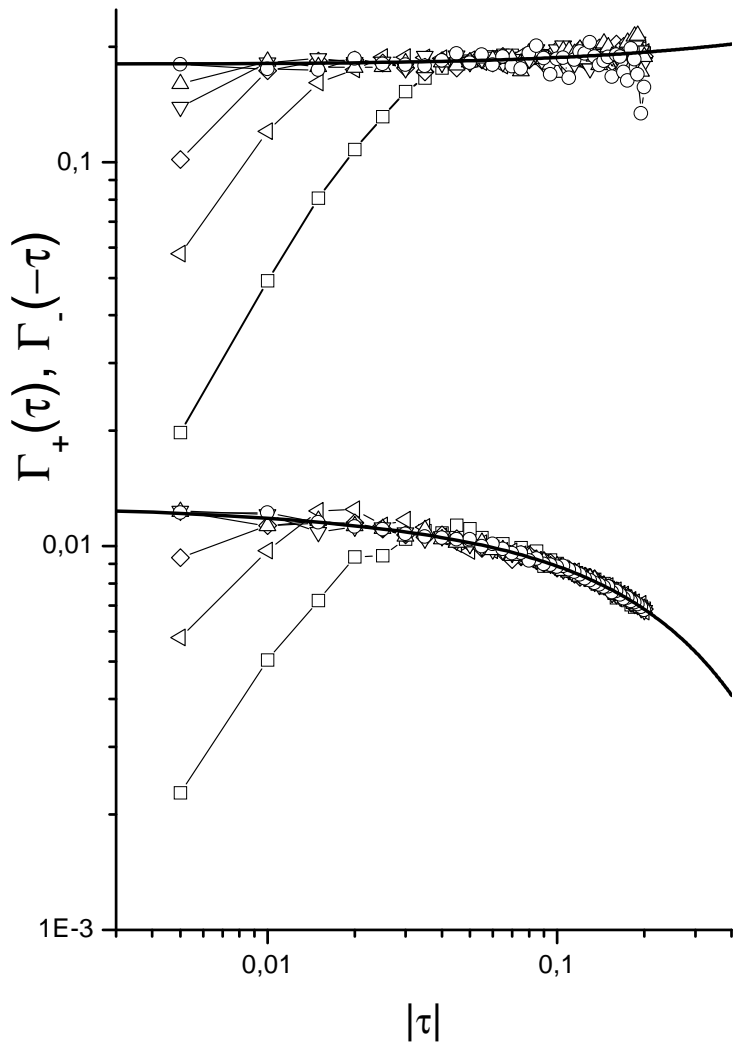,width=12cm} 
 
\caption{Finite-size behaviour of the effective amplitude of the 
susceptibility data in the critical region window. The data for lattice 
sizes $L=20,40,60,80,100$ and $200$ are denoted by squares, 
left-triangles, diamonds, down-triangles, up-triangles and circles, 
respectively. The high-temperature data correspond to the upper part and 
the low-temperature data to the lower part of the figure. The data 
generated from the high- and low-temperature series expansions are shown 
by thick solid lines.} 
 
\label{fig1} 
\end{figure}

The MC results are shown in the Figure~\ref{fig1} together with the series 
expansion data which are denoted by thick solid lines. The series data 
compare well with the MC data also for size $L< 80$ when $|\tau|>0.01$. 
The corrections to scaling are not small in the interval of the reduced 
temperature accessible by the Monte Carlo simulations. 
 
\subsection{Fit to the data} 
 
We have fitted the data taken  
in the critical window of the LT and HT phases to 
the following expression 
 
\begin{equation} 
\chi=\Gamma_\pm |\tau|^{-\gamma}(1+a_\pm|\tau|^{\Delta})+D_\pm 
\label{chi-fit} 
\end{equation}  
 
\noindent where $\gamma=13/9$ and $\Delta=y_{T,2}/y_{T,1}=2/3$ are known 
exactly~\cite{N}, The value of $\Delta$ is supported by the series 
expansion analysis of Ref.~\cite{AP}, by a finite-size 
analysis~\cite{vGRV} of the 3-state Potts quantum chain, and by a 
finite-size analysis of the transfer matrix results of Ref.~\cite{dQ}.   
The constant ``background'' terms $D_\pm$ are known to be important even 
for the Ising model~\cite{SV}, especially so in the LT phase. We have kept 
a single correction-to-scaling term $a_{\pm}|\tau|^{\Delta}$ in order to 
avoid introducing too many fitting parameters. However, we have also 
checked the stability under inclusion of further terms, e.g., 
$(1+a_\pm|\tau|^{\Delta}+b_\pm|\tau|+c_\pm|\tau|^{2\Delta})$ and found 
that $\Gamma_\pm$ varies only within the accuracy of the fit. 
 
The results of the fit to the susceptibility data are shown in 
Table~\ref{table1} for lattices sizes $L=20,40,60,80,100$ and $200$. The 
right-hand boundary of the critical window was chosen as 
$|\tau|=|(T-T_c)/T|=0.2$ and the left-hand boundary according to the 
lattice size as $|\tau|>1/L^{1/\nu}$. 
 
\begin{table} 
\caption{Results of the fit to the MC susceptibility data  
in the critical region window. For comparison, we have also reported in the  
last line the results of a similar fit to the series expansion (SE) data.} 
\center 
\begin{tabular}{|r|l|l|l|l|l|l|l|} 
$L$ & $\Gamma_-$ &  $a_-$   & $D_-$     & $\Gamma_+$&  $a_+$   & $D_+$ &  
$\Gamma_+/\Gamma_-$ \\ \hline 
20  & 0.01321(3) & -1.61(1) & 0.0087(5) & 0.2036(6) & -0.67(2) & 0.37(1) & 
15.41(9) \\ 
40  & 0.01260(2) & -1.500(9)& 0.0071(3) & 0.1966(3) & -0.64(1) & 
0.448(8) & 15.60(5) \\ 
60  & 0.01245(4) & -1.375(1)& 0.0026(1) & 0.1642(3) & 0.90(2)  & 
-0.228(6) & 13.19(7) \\ 
80  & 0.01251(1) & -1.393(7)& 0.0026(3) & 0.1815(2) & 0.14(1)  & 
0.008(5) & 14.51(3) \\ 
100 & 0.01250(1) & -1.41(6) & 0.0037(3) & 0.1728(16)& 0.49(9)  & 
-0.11(5) & 13.82(16) \\ 
200 & 0.01273(1) & -1.509(6)& 0.0070(2) & 0.1741(2) & 0.46(1)  & 
-0.17(1) & 13.68(3) \\ \hline 
SE  & 0.012774(3)& -1.517(8)& 0.0070(2) & 0.1783(7) & 0.24(2)  & 
0.005(6) & 13.96(7) 
\end{tabular} 
\label{table1} 
\end{table} 
 
 For comparison, we have also fitted in the same way the series expansion
data, taking as critical window the interval $|\tau|\in [0.01:0.2]$ and
assuming that the series expansion data become increasingly accurate for
larger values of $|\tau|$.  The error bars reflect the accuracy of the fit
and do not allow for systematic deviations due to the fact that the
critical window is far from the asymptotic limit when large corrections to
scaling are present.  The fit to the MC data shows stability and
consistency with the series expansion data, listed in the last line of
Table~\ref{table1}.

Very conservatively, we can conclude from our MC data and Pad\`e
approximation of series expansion that the ratio of the susceptibility
amplitudes for the 3-state Potts model is $\Gamma_+/\Gamma_-=14\pm1$.  
Thus, our results are completely consistent with the value
$\Gamma_+/\Gamma_-=13.848$ calculated by Delfino and Cardy~\cite{DC}.
 
An additional check of the results is obtained by studying the ratio 
$r^{(SE)}(\tau)$ of the HT and the LT effective amplitudes  
of the susceptibility, as computed 
from series expansions  
$r^{(SE)}(|\tau|)=\Gamma_+(\tau)/\Gamma_-(-\tau)$. We can expect the 
following behaviour of this ratio 
 
\begin{equation} 
r^{(SE)}(|\tau|)=\frac{\Gamma_+}{\Gamma_-}\left(  
1+a_1|\tau|^{\Delta}+a_2|\tau|+a_3|\tau|^{2\Delta}+... 
\right) 
\label{chi-fit-rat}    
\end{equation} 
 
\noindent as $\tau \rightarrow 0+$. 
 
\noindent A three-parameter ($\frac{\Gamma_+}{\Gamma_-}$, $a_1$, and 
$a_2$)  fit of $r^{(SE)}(|\tau|)$ in the temperature window $[0.01:0.1]$ 
gives $\frac{\Gamma_+}{\Gamma_-}=14.2(5)$. 
 
In the case of the Ising model ($q=2$ Potts model), a similar fit of the 
same ratio $r^{(SE)}(|\tau|)$  to the form 
 
\begin{equation} 
r^{(SE)}(|\tau|)=\frac{\Gamma_+}{\Gamma_-}\left( 
1+a_1|\tau|+a_2|\tau| \log |\tau| 
\right) 
\label{chi-fit-ising} 
\end{equation} 
 
\noindent leads to a value of the critical amplitude ratio
$\frac{\Gamma_+}{\Gamma_-}=39\pm 2$ which is consistent with the exact
value $37.69365...$. The large error bars are due to the fact that we have
used series expansions of the same length as for the 3-state Potts model
in order to test under similar conditions the accuracy of the method.  Of
course, much better results could obtained fully using the much longer
series expansions which are available for the 2d Ising model~\cite{ONGP}.
 
\section{Summary of the results and conclusions} 
\label{Summary} 
 
The values of the suceptibility critical amplitude ratio for the Potts
model with $q= 2,3$ and $4$ were calculated by Delfino and Cardy~\cite{DC}
using the two-kink approximation of the exact scattering theory for the
Potts model~\cite{CZ}. The value $37.699$ thus obtained for the ratio in
the $q=2$ case (Ising model) agrees well with the exactly known value.  
However the same authors and Barkema were unable to confirm
numerically~\cite{DBC} their theoretical results: $13.848$ for the $q=3$
Potts model and $4.013$ for the 4-state Potts model. The discussion of the
latter case is beyond present paper. However by analysing our MC data and
the existing series expansions, we find that the critical amplitude ratio
for the 3-state Potts model can be very safely identified with the
estimate $14\pm 1$, quite consistently with the prediction by Delfino and
Cardy.
 
What is the main difference between the analysis of Ref.~\cite{DBC} and that
presented here? First, we calculate the amplitudes separately in both the 
LT and the
HT phases by fitting the temperature behaviour of the susceptibilities. It is
also important that the value of susceptibility was computed by not less
than $10^5$ Wolff MonteCarlo steps at each value of temperature. Indeed the 
fit to the susceptibility becomes unstable for  smaller statistics. Next, we
have computed the amplitude ratio as a function of temperature defined in 
 the same way in both phases. 
This is not the case for the analysis in Ref.~\cite{DBC}, where
the corresponding  temperatures in the two phases are shifted 
by a factor proportional to the ratio of the correlation lengths.

Since two out of the three $\Gamma_+/\Gamma_-$ ratio values computed by
Delfino and Cardy agree well, either with the known exact result for the
Ising case, or with our MC and series expansion data for $q=3$, little
doubt remains, in our opinion, that also their prediction for the 4-state
Potts model may be correct. We wish to quote here a MC analysis of the
4-state Potts model by Caselle et al.  where an estimate consistent with
the Delfino and Cardy prediction is obtained. However Delfino et
al.~\cite{DBC} did not found this analysis completely satisfactory and
therefore the susceptibility ratio prediction for $q=4$ is still waiting
for further numerical verifications.
 
\section{Acknowledgements} 
 
LNS is grateful to the Theoretical Physics group of the Milano-Bicocca
University and to the Statistical Physics group of the University Henri
Poincar\'e Nancy-I for their kind hospitality. Financial support from the
twin research program between the Landau Institute and the Ecole Normale
Sup\'erieure de Paris as well as financial support from the CARIPLO
Foundation and Landau Network---Centro Volta, and Russian Foundation for
Basic Research under project 99-07-18412 are also acknowledged.

\end{document}